# Weak Localization and Magnetoconductance in Percolative Superconducting Aluminum Films

Kazumasa Yamada[a], Bunjyu Shinozaki, and Takashi Kawaguti

Department of Physics, Kyushu University, 4-2-1 Ropponmatsu, Chuo-ku, Fukuoka,

810-8560, Japan

In order to investigate the crossover from the homogeneous behavior to inhomogeneous (percolative) one, the temperature $T$ and magnetic field $H$ dependence of the sheet resistance $R_\square$ have been measured for two-dimensional granular aluminum films. Fitting the theory to data of magnetoconductance near $T_C$ with use of the diffusion constant $D(T)$ as a fitting parameter, we have obtained the anomalous $T$-dependent diffusion constant $D$. From the analysis of $D(T)$, the electron diffusion index θ, a certain critical exponent in percolation theory, has been obtained. In the relation $R_\square$ - θ, the value of θ varies abruptly near 1.5kΩ. This behavior suggesting the above mentioned crossover is similar to our previous results determined from the temperature dependence of the upper critical field. For percolative films in $H$=5T, we have found the strong $R_\square$ dependence of the prefactor $\alpha_T$ in the expression $\sigma = \left[\alpha_\mathrm{T} e^2 / \left(2\pi^2 \hbar\right)\right] \ln T + \sigma_0$. The relation $\alpha_T \propto 1/R_\square$ can be explained qualitatively by a model of scaling law for percolation.

[a] Present address: Tarucha Mesoscopic Correlation Project, ERATO, JST, 4S-308S, NTT Atsugi Research and Development Center, 3-1 Wakamiya, Morinosato, Atsugi 243-0198.

E-mail:yamada@tarucha.jst.go.jp

## 1. INTRODUCTION

Superconductor-insulator mixtures and granular superconductors have been widely investigated, in order to clarify the interplay among percolation, Anderson localization and superconductivity. For homogenous dirty superconductors, superconducting correlation length $\xi_S$ is simply given by $\xi_S^2 = D\tau_{GL}$, where $D$ and $\tau_{GL}$ are the diffusion constant and the Gintzburg-Landau relaxation time $\tau_{GL} = \pi\hbar/\{8k_B|T_C - T|\}$, respectively. However, electron diffusion in percolative superconductors is different from that in homogeneous ones. Here, percolative superconductors can be characterized by the condition $\xi_P > \bar{\xi}_S$, where $\xi_P$ and $\bar{\xi}_S$ are the percolation correlation length and the effective superconducting correlation length, respectively. It is known that the diffusion constant $D$ has an anomalous scale-dependence $D(L) \propto L^{-\theta}$, where $\theta \approx 0.9$ is the diffusion index[1,2]. From the relation $\bar{\xi}_S^2 = D(\bar{\xi}_S)\tau_{GL}$, we obtain $\bar{\xi}_S \propto \tau_{GL}^{1/(2+\theta)} \propto |T_C - T|^{-1/(2+\theta)}$. As a result of anomalous $T$-dependent $D$, the value of the upper critical magnetic field $H_{C2}$ near $T_C$ for the case of $\xi_P > \bar{\xi}_S$ is expected to show the anomalous temperature dependence[ 3 ], $H_{C2} \propto 1/\bar{\xi}_S^2 = 1/(D(\bar{\xi}_S)\tau_{GL}) = (1 - T/T_C)^{2/(2+\theta)}$. This power-law relation between $H_{C2}$ and $T$ at temperatures $T<T_C$ near $T_C$ had been observed in three-dimensional (3D) In-Ge, Al-Ge[4] and 2D Pb[5].

We have investigated $H_{C2}(T)$ in 2D granular Al films in the previous work[6]. The index $\theta$ increases steeply near characteristic sheet resistance $R^* \approx 1.5\mathrm{k}\Omega$ and approaches $\approx 0.9$ with increase of $R_\square$. In films with $R_\square > R^*$, the excess conductance $\sigma'$ due to superconducting fluctuations decreases also drastically as compared with that estimated from theories for homogenous films. These increment and decrement can be explained qualitatively by the model based on the scaling law of percolation. We will extend the idea of the percolation model, in order to examine the effects on weak localization and the magnetoconductance $\Delta\sigma$ at $T>T_C$.

The formula for $\Delta\sigma$ in homogenous dirty-superconductors is given by the theory including the diffusion constant $D$. Therefore, the temperature dependence of $D$ can be also directly investigated by the analysis of $\Delta\sigma$ near $T_C$. We expect that $D(T)$ shows the same anomalous temperature dependence as that determined from the analysis of $H_{C2}(T)$ because of the symmetrical temperature dependence of $\tau_{GL}(T)$ with respect to $T_C$.

As the temperature decreases, the interference of a single electron wave scattered by random potentials is enhanced and brings the precursor effect of $\ln T$ dependence of the sheet conductance as $\sigma = [\alpha_T e^2/(2\pi^2\hbar)]\ln T + \sigma_0$. It has been reported that the weak localization effect in percolative films is weaker than that expected from the theory for homogenous films[7, 8]. In the networks with loops larger than the phase relaxation length $L_{in}$, the interference effect is suppressed. It is considered that $\alpha_T$ decreases with increasing $\xi_P$ beyond $L_{in}$ [9,10]. However, disorder dependence of the prefactor $\alpha_T$ has not been studied well.

In this paper, we report the experimental data for crossovers from the homogeneous behavior to a percolative one in the diffusion constant and electron localization effect. From the experimental data on $D(T)$ determined by the analysis of the magnetoconductance near $T_C$, we will discuss the $R_\square$ dependence of diffusion index $\theta$.

As a reason for good correlation of superconducting properties with $R_\square$, we will discuss the effect of quantum percolation in the present superconducting granular films.

## 2. THEORETICAL BACKGROUND

At first, theories for homogenous 2D films are explained. Secondly, we will show theories for percolative films.

According to theories for weak electron-localization[8] and Coulomb interaction[7], temperature dependence of conductance due to quantum corrections in normal metallic films is given by,

$$\sigma' = [(1-F)+p]\frac{e^2}{2\pi^2\hbar}\ln T = (\alpha_{T,I}+\alpha_{T,L})\frac{e^2}{2\pi^2\hbar}\ln T = \alpha_T \frac{e^2}{2\pi^2\hbar}\ln T. \quad (1)$$

Here $(1-F)$ is a screening term which goes to zero ($F=1$) in the limit for the short range interaction and goes to 1 ($F=0$) in the limit for the long range interaction; $p$ is the exponent of temperature term in the inelastic scattering rate $1/\tau_{in} \propto T^P$. Even in superconducting films, $\sigma'$ can be expressed by Eq.(1) under high magnetic fields which can suppress superconductivity. In the low temperature region, inelastic scattering is mainly due to electron-electron(e-e) processes. For the films whose thermal length $\tau_T$ is longer than the film thickness $\left(\tau_T = \sqrt{\hbar D/k_B T} > d\right)$, the inelastic scattering rate due to e-e processes has been calculated by Al'tschuler $et\ al.$[11] and Fukuyama and Abrahams[12] as follows,

$$1/\tau_{ee} = \frac{k_B T}{\hbar}\frac{e^2}{2\pi^2\hbar}R_\square \ln\left(\frac{\pi\hbar}{e^2 R_\square}\right). \quad (2)$$

There are two contributions of superconducting fluctuations to the conductance of thin films above $T_C$. The first is the Aslamazov-Larkin contribution $\sigma'_{AL} = e^2/(16\hbar\eta)$, where $\eta = \ln(T/T_C)$ [13]. The second is the Maki-Thompson contribution $\sigma'_{MT}$ [14, 15]. The improved expression by Reizer [16] is given by

$\sigma'_{MT} = [e^2/(2\pi\hbar b_2^{1/2})][\ln(1+B_2) + 2B_2 \ln(1+1/B_2)]$, where $b_2 = [4e^2/(3\pi\hbar)]R_N$ and $B_2 = \pi b_2^{1/2}/(8\eta)$.

The value of $R_N$ is the sheet resistance at the normal state. Thus, the temperature dependence of $R_\square$ in 2D superconductors is given by,

$$1/R_\square = 1/R_N + \sigma'_{AL}(T,T_C) + \sigma'_{MT}(T,T_C,R_N) + [\alpha_T e^2/(2\pi^2\hbar)]\ln(T/T_0). \quad (3)$$

Magnetoconductance $\Delta\sigma$ is defined as $\Delta\sigma(T,H) = \sigma(T,H) - \sigma(T,0)$. The value of $\Delta\sigma$ in 2D dirty superconductors is given by superconducting fluctuations AL and MT terms and weak localization; $\Delta\sigma(T,H) = \Delta\sigma_{AL} + \Delta\sigma_{MT} + \Delta\sigma_L$. The $\Delta\sigma_{AL}$ was calculated by Abraham and Tsuneto[17] as follows,

$$\Delta\sigma_{AL}(H,T) = -\frac{e^2}{16\hbar\eta}(1 - g_{AL}), \quad (4)$$

where $g_{AL} = 2\left(\frac{2\eta}{\lambda_0 a}\right)^2 \left[\frac{\lambda_0 a}{2\eta} + \Psi\left(\frac{1}{2} + \frac{\eta}{\lambda_0 a}\right) - \Psi\left(1 + \frac{\eta}{\lambda_0 a}\right)\right]$, $\Psi$ is the Digamma function, $\lambda_0 = \frac{\pi\hbar}{8k_B T}$ and $a = \frac{4eDH}{c\hbar}$.

The $\Delta\sigma_{MT}$ was calculated by Lopes dos Santos and Abrahams[18] as follows,

$$\Delta\sigma_{MT} = -\frac{e^2}{2\pi^2\hbar}\beta(\eta,\tau_{in})\left[Y(\tau_{in}a) - Y\left(\frac{\lambda_0}{\eta}a\right)\right], \quad (5)$$

where $Y(x)$ is given by $Y(x) = \ln(x) + \Psi(x)$. When $\eta \ll 1$, $\beta(\eta,\tau_{in})$ is given by $\beta(\eta,\tau_{in}) = (\pi^2/4)[\eta - \pi\hbar/(8k_B T\tau_{in})]$. The $\Delta\sigma_L$ is shown as follows[19],

$$\Delta\sigma_L = \frac{e^2}{2\pi^2\hbar}\left[\frac{3}{2}Y(\tau_1 a) - \frac{1}{2}Y(\tau_{in}a)\right], \quad (6)$$

where $1/\tau_1 = 1/\tau_{in} + 4/3 \cdot 1/\tau_{SO}$.

The electron diffusion problem in percolative specimens is treated as random walk on the random network[1]. We consider first a particle which has started on the infinite cluster, at one point 0, and moves during time $t$ by a distance $r(t)$. In macroscopic limit, when $r \gg \xi_P$, $\langle r^2(t) \rangle = 2Dt$ is expected, where $D$ is a certain diffusion constant for a particle. On the other hand, if $a < r(t) < \xi_P$ is satisfied, then self-similarity implies that $\langle r^2(t) \rangle \propto t^{2/(2+\theta)}$, where $a$ is a length of the order of the typical grain size. When a scale dependent diffusion coefficient is defined as $\langle r^2(t) \rangle = D(r)t$, the relation $D(r) \propto r^{-\theta}$ is satisfied at the condition $a < r < \xi_P$. This equation will be the starting point of our following discussion on granular superconductors.

For percolative superconducting films, the value of $D$ in the expression for $\Delta\sigma$ due to superconducting fluctuations above $T_C$ is not constant but determined on the length range $\bar{\xi}_S$ as mentioned in the introduction. Therefore, we can consider that the value of $D(\bar{\xi}_S)$ depends on $T$ in the region $\bar{\xi}_S < \xi_P$ as follows,

$$D(\bar{\xi}_S) \propto \tau_{GL}^{-\frac{\theta}{2+\theta}} \propto |1 - T/T_C|^{\frac{\theta}{2+\theta}}. \qquad (7)$$

A model for weak localization phenomena in 2D percolation networks was proposed by Palevski and Deutscher[9]. Within the framework of their model, the percolation network can be divided into two basic parts, if $\xi_P \gg (L_{in}, w)$ ($w$ indicates the width of the channel). One part, on the scale of $L_{in} \cong [D(L_{in})\tau_{in}]^{1/2}$, consists of many loops of different radii smaller than that of the order of $L_{in}$. The other is the rest of the network which on this scale does not contain loops smaller than $L_{in}$. The first part can be regarded as a 2D system with an average sheet resistance $R(L_{in})$ on the scale $L_{in}$. Since the second term becomes negligible, they obtained following relation

$$\alpha_{T,L}^{inhomo.} = \frac{R(L_{in})}{R_\square} \alpha_{T,L}^{homo.}, \tag{8}$$

where $\alpha_{T,L}^{homo.}$ and $\alpha_{T,L}^{inhomo.}$ are the coefficient of $\ln(T)$ term in Eq.(1) due to weak localization in homogenous and inhomogeneous films, respectively. It can be considered that the measured sheet resistance $R_\square$ of a high resistive film in the percolative region is determined mainly by the change of a coupling strength between grains and depends strongly on the strength. Therefore, It is considered that the length $L_{in}$ can be regarded as almost constant for high resistance films and we obtain following relation,

$$\frac{R(L_{in})}{R_\square} = \begin{cases} 1 & (\xi_P < L_{in}) \\ \propto 1/R_\square & (\xi_P \gg L_{in}) \end{cases}. \tag{9}$$

According to percolation theory, the resistance of discontinuous films varies as $R \propto (p - p_C)^{-\mu}$, $p(>p_C) \to p_C$, where $p$, $p_C$ and $\mu$ are the film coverage, the critical coverage and the critical conductance exponent, respectively. Since the percolation correlation length diverges as $\xi_P \propto |p - p_C|^{-\nu}$, the variation of the DC resistance can be expressed in terms of $\xi_P$ as $\xi_P \propto R^{\nu/\mu}$.

## 3. SAMPLE PREPARETIONS

Aluminum films were made by deposition onto glass substrates patterned by photolithography. We prepared films in wide ranges of $d$ and $R_\square$ by two different methods.

(1) single-layer method; Aluminum was evaporated in a pressure $1\times10^{-5} \sim 3\times10^{-4}$ mb with a deposition rate 0.5 Å/s~1 Å/s. Except for one 190 Å thick film, the range of thickness is from 60 to 100 Å.

(2) multi-layer method (almost the same method as that for FOMP[20])); After deposition of 30~40 Å Al film in a pressure $1\times10^{-5} \sim 3\times10^{-5}$ mb with a deposition rate ~2 Å /s, an Al film was oxidized in $2.0\times10^{-4}$ mb for 1 minute. We repeated this process 4 times. The range of total thickness is from 120 to 180 Å. By this method, we can prepare sufficiently thick but high resistive films.

We connected a personal computer with the thickness monitor of quartz crystal oscillator and the digital voltmeter. We recorded *in situ* the resistance $R(d)$ during deposition. It has been found that the relation $R(d) \propto (d-d_C)^{-\mu}, \mu = 1.2 \sim 1.3$ is satisfied in the range $d - d_C < 30 \text{Å}$, which is explained by the percolation model. The value of $d_C$ depends on deposing conditions. We have mentioned details of measurement method and $d$ dependence of $R_\square$[6]. According to the TEM micrograph of thin Al-Al$_2$O$_3$ film taken by Laibowitz *et al.*[21], clusters of Al-Al$_2$O$_3$ films are the labyrinth-like structure.

## 4. EXPERIMNTAL RESULTS AND DISCUSSIONS

### A. Magnetoconductance above $T_C$

The clusters in percolative films are discontinuous. The only maximum cluster expands from the end to the end. The maximum cluster has multi-fractal geometry. Because the dimension of the maximum is intermediate value between 1 and 2 in the scale $(L < \xi_P)$, the electron diffusion is weaker than that of two-dimensional film. As $T$ approaches $T_C$, extending of $\bar{\xi}_s$ is shorter than that of homogenous 2D and $D(\bar{\xi}_S)$ is suppressed. It is considered that $D(\bar{\xi}_S)$ above $T_C$ depends on $T$ similarly to that below $T_C$.

We measured the resistance of films above $T_C$ in magnetic fields $H \leq 5\text{T}$. For analyses of $\Delta\sigma$ with the sum of quantum correction terms due to fluctuation and weak localization effects, we fit first Eq.( 3 ) to data of $R_\square(T)$ using $\alpha_T$ and $T_C$ as fitting parameters. Secondly, we analyzed $\Delta\sigma$ using this value of $T_C$. We fit the sum of Eqs.( 4 )-( 6 ) to data with using $D$ as a fitting parameter, where the inelastic scattering time $\tau_{in}$ is calculated from Eq.( 2 ). Figure 1 shows the $H$ dependence of magnetoconductance at various temperatures for a typical inhomogeneous film. The lines show the theoretical ones obtained from the above procedure. At low magnetic fields, the relation $\Delta\sigma \propto H^2$ is satisfied as expected from fluctuation theories.

Figures 2 shows temperature dependence of $D$ for films near the crossover region. In relatively high temperature regions, the value of $D$ decreases as $T$ decreases. We fit Eq.( 7 ) to data at the relatively high temperate region, regarding $\theta$ as a fitting parameter. For the film with the lowest resistance, the value of $D$ is almost constant in a large temperature region near $T_C$. This behavior is the same as that of homogenous two-dimensional films. For films with larger value of $R_\square$, it is clear that the crossover occurs from the region with $T$-independent $D$ to that with $T$-dependent $D$. This crossover suggests that the inhomogeneous behavior ($\bar{\xi}_S < \xi_P$) changes to the homogenous behavior ($\bar{\xi}_S > \xi_P$) as $T$ approaches to $T_C$ and the length $\bar{\xi}_S$ increases. Although the temperature corresponding to each crossover does not vary systematically with increase of $R_\square$, such crossover behavior is reasonable because the length $\xi_P$ increases and the temperature region with $\bar{\xi}_S > \xi_P$ decreases as the resistance $R_\square$ increases.

On the other hand, the value of $D$ below $T_C$ is obtained from data of $H_{C2}$ using the relation $D = 4ck_B/(\pi e)[H_{C2}(T)/(T_C - T)]^{-1}$. In the $R$-$T$ curve at a constant $H$, $T_C(H)$ was defined as a temperature at which half of $R_N$ was restored[6]: We had noted that we obtain the similar power law $H_{C2}$-$T$ relation for percolative films by the use of different criterions $T_C(H)$. Strictly speaking, however, $H_{C2}$-$T$ relation, namely, temperature dependence of $H_{C2}$ near $T_C$ seems to depend on this criterion. This means that there is some ambiguity in the data of $D(T)$ at especially very near $T_C$. In fact, we could not observe evident crossover as shown in Fig.2 in $H_{C2}$-$T$ analysis. On the contrary, for the analysis of magnetoconductance, we can obtain reliable temperature dependence of $D$ by the use of the only value of $T_C$ at zero magnetic field. Therefore, the $D(T)$ in Fig.2 indicates the crossover from inhomogeneous region to homogeneous one.

Figure 3 shows the $R_\square$ dependence of $\theta$. The values of $\theta$ are nearly 0 for films with smaller than 1.5 k$\Omega$. This means the scale-independent $D$ for films in the homogenous region, where the condition $\bar{\xi}_S > \xi_P$ is satisfied. As $R_\square$ increases, index $\theta$ deviates near 1~2 k$\Omega$ from the solid line $\theta = 0$ for the homogeneous system. The values of $\theta$ in the large $R_\square$ region are close to that estimated from the classical percolation model as shown by dotted line. The $R_\square$ dependence of index $\theta$ from the analysis of $\Delta\sigma$ is almost the same as that obtained from the analysis of $H_{C2}(T)$ shown by different marks.

## B. Electron weak localization

Figure 4 shows temperature dependence of the sheet conductance $\sigma$ for various multi-layer films in a magnetic field 5 T. Behaviors of single-layer films are discussed in the following section. The conductance of the present films shows the $\ln T$ dependence and superconductivity seems to disappear at measured temperatures. The data with the lowest value of conductance show some deviation from the $\ln T$ dependence. However, it cannot be considered that these films are in strong electron localization regime, because the temperature dependence does not show the $R \propto e^{1/k_B T}$ expected for the strong localization. By fitting Eq.(1) to data of $\sigma$ vs $\ln T$, we determined the coefficient $\alpha_T$.

Figure 5 shows the $R_\square$ dependence of $\alpha_T$. The value of $\alpha_T$ is almost the constant value $\approx 2$ in relatively clean films. This value is reasonable, if we take account of the sum of two contributions of weak localization ($p=1$) with inelastic scattering rate due to e-e scattering and Coulomb anomaly ($F=0$) in Eq. (1). The value of $\alpha_T$ decreases when $R_\square$ increases beyond 6~8 k$\Omega$. This decrease cannot be explained by the theory of weak localization and Coulomb anomaly for 2D homogenous system. We consider that this behavior is due to crossover from homogenous to inhomogeneous films. For the purpose of explanation of the $R$ dependence of $\alpha_T$, we apply the percolation model[6] for weak localization phenomena in 2D percolation networks to the present data. The fact that the value of $\alpha_T$ is below unity indicates that percolation reduces not only weak localization but also Coulomb anomaly effects. However, there is no theoretical prediction for the percolation effect on the Coulomb anomaly. Assuming that the effect on the coefficient $(1-F)$ in Eq.( 1 ) due to Coulomb anomaly is similar to weak localization effects, we obtain the relation $\alpha_T \propto 1/R$ $(\xi_P \gg L_{in})$ from Eq.( 8 ), which is shown by the solid line in Fig. 5. For clarifying the crossover in $\alpha_T - R_\square$ relation and also the effect of percolation on the Coulomb anomaly, not only theoretical investigation but also detailed experimental studies of temperature and magnetic field dependence of dirty films are necessary.

## C. The $R_\square$ and $d$ dependence of transport properties

We mentioned that the crossover from homogenous to inhomogeneous films occurs as $R_\square$ increases, where we assumed the following relation $\xi_P \propto R_\square^{\nu/\mu}$. We prepared the films whose $R_\square$ was almost the same but thickness was different. In Figs. 6(a), (b) and (a'), (b'), we show $d$ and $R_\square$ dependence of $\theta$ and $\alpha_T$, in order to clear the relevant scaling parameter. For data of $\theta$, we showed the result from the analysis of $H_{C2}(T) = H_{C2}{}^0 (1 - T/T_C)^{2/(2+\theta)}$ at temperatures below $T_C$. The horizontal axis $R_\square$ means $R_N$, which is obtained from the best fit of Eq.( 3 ) to the data of $R_\square(T)$. The closed and open marks show data of the multi-layer films whose thickness is thicker than 100Å and data of the single-layer films whose thickness is thinner than 100Å, respectively. By single layer method, we obtained exceptionally a thick film whose sheet resistance is relatively high: The data are shown by triangular marks(▲). The mark (■) shows the data of films that consist of 10 layers.

Figures 6(a), (b) show $d$ dependence of these values. From the data of single-layer films, the values of $\theta$, $\alpha_T$ seem to correlate with $d$. If it is assumed that the coverage $p$ is relates to $d$, these behaviors are reasonable. However, the data of the thick single-layer

film and multi-layer films deviate from these behaviors. It means that these values do not depend on $d$. On the other hands, as known from Fig.6(a'), it is found that there is no remarkable difference in $\theta - R_\square$ relation among multi-layer thick films, single-layer thin films and even exceptional thick single-layer film. The superconducting properties, not only $\theta$ but also $T_C$ and $H_{C2}{}^0$ in the expression of $H_{C2}(T) = H_{C2}{}^0 (1 - T/T_C)^{2/(2+\theta)}$, show good correlation with $R_\square$ [6].

As a reason for correlations of superconducting properties with $R_\square$, following quantum percolation model may be considered. We regard these films as random networks which consist of grains bound by junctions whose normal tunneling resistance $R_N{}^{micro.}$ varies in place to place. Theories point out that in such a junction with $R_N{}^{micro.}$, the phase coherence between two constituent superconductors at $T=0$ can establish only when $R_N$ is smaller than $h/4e^2$. When such circumstances can be allowed at finite temperatures, it can be considered that there are both Josephson junction and non-Josephson junction in a film and that the effective percolation coverage, that is, the effective $\xi_P$ in *superconducting* films depends on the ratio of the number of Josephson junctions to that of all. Taking into account of that this ratio depends on the macroscopic $R_\square$, it is reasonable that superconducting parameters correlate to the macroscopic $R_\square$. For further investigations of the quantum percolation, not only experimental but also theoretical detailed studies are necessary.

Figure 6(b') shows $R_\square$ dependence of $\alpha_T$. It is found that there is a remarkable difference of $\alpha_T$ between two kinds of films. The value of $\alpha_T$ in single-layer films seems to decrease when the $R_\square$ increases beyond 1~2kΩ. Thick single layer films show the same behavior as that of multi-layer films. It seems that the crossover resistance of single-layer thin films is smaller than that of multi-layer thick films. It is deduced in framework of a classical percolation model that $\alpha_T$ is a function of $d$ because $\xi_P$ is longer in thinner films. However, Figs.6(b) and (b') show that $\alpha_T$ is a function of neither thickness $d$ nor sheet resistance $R_\square$. In order to examine whether $\alpha_T$ depends on the number of layers, we made the films which consist of 10 layers. The data are shown by closed square mark in Figs.6(b) and (b'). It is shown that $\alpha_T$ is not determined by the number of layers. At the present stage, we have no explanation for this difference of $\alpha_T$ between single-layer and multi-layer films.

## 5. CONCLUSION

For clarifying the crossover from homogeneous to percolative behaviors, we have investigated the superconducting and electron localization properties of evaporated granular aluminum films. In order to change the coupling strength between grains and also the area coverage, we have adopted two different kinds of film preparation. One of them is single thin layer method and the other is multi thick layer method. We have analyzed the data of $R_\square$ ($T,H$) as follows; (1) $\Delta\sigma = \Delta\sigma_{AL} + \Delta\sigma_{MT} + \Delta\sigma_L$ for the magnetoconductance. (2) $\sigma = \alpha_T \left( e^2 / 2\pi^2 \hbar \right) \ln T$ for the conductance in a high magnetic field $H$=5T.

It has been found that the superconducting properties θ, $T_C$ and $H_{C2}^0$ correlate much better with the sheet resistance $R_\square$ than the films thickness $d$. This result indicates that the percolation length $\xi_P$ is determined by $R_\square$, but not by $d$. For this reason, we discussed the quantum percolation model.

In order to confirm the $T$ dependence of diffusion constant $D(T)$ above $T_C$, we fit the theories to the data of $\Delta\sigma$ with use of $D(T)$ as a fitting parameter. The value of $D(T)$ decreases as $T$ approaches $T_C$. This result is consistent with the behavior expected from the percolation model. Because the value of $D(T)$ slightly depends on the value of $T_C$, we cannot refer the quantitative comparison of the strength of $D(T)$ between results from $\Delta\sigma$ analysis above $T_C$ and $H_{C2}$ analysis below $T_C$.

Although the value of coefficient $\alpha_T$ is almost constant $\approx 2$ in the homogeneous region for both single and multi films, the $\alpha_T$ for multi films decreases steeply with increase of $R_\square$ and shows $\alpha_T \propto 1/R_\square$ in the region beyond $R_\square \cong 5-8 k\Omega$. For the $R_\square$ dependence of $\alpha_T$, however, there is a large difference of $\alpha_T$ between thin single films and thick multi films. At present, we have no exact explanation for this difference.

[1] P. G. de Gennes, in *NATO ASI series. Series B; physics* v.109 p.83 (Plenum Press, New York and London, 1984) Proceedings of the NATO Advanced Study Institute on Percolation, Localization and Superconductivity, edited by A. M. Goldman and S. A. Wolf (Les Arcs, Savoie, France, 1983).

[2] D.Stauffer and A.Ahanory, *Introduction to percolation theory* (Taylor and Francis, London, 2nd edition 1994).

[3] S. Alexander and E. Halvy, J. Phys. (France) **44**, 805 (1983).

[4] A. Gerber and G. Deutscher, Phys. Rev. B **35**, 3214 (1987).

[5] A. Gerber and G. Deutscher, Phys. Rev. Lett. **63**, 1184 (1989).

[6] K. Yamada, H. Fujiki, B. Shinozaki, and T. Kawaguti, Physica C **355**, 147 (2001).

[7] B. L. Altshler, A. G. Arnov, and P. A. Lee, Phys. Rev. Lett. **44**, 1288 (1980).

[8] E. Abrahams, P. W. Anderson, D. C. Licciardello, and T. V. Ramakrishnan, Phys. Rev. Lett. **42**, 673 (1979).

[9] A. Palevski and G. Deutscher, Phys. Rev. B **34**, 431 (1986).

[10] M. Aprili, J. Leseur, L. Dumouin, and P. Nedellec, Solid State Commun. **102**, 221 (1997).

[11] B. L. Altshuler and A. G. Aronov, Solid State Commun. **30**, 115 (1979).

[12] H. Fukuyama and E. Abrahams, Phys. Rev. B **27**, 5967 (1983).

[13] L. G. Aslamazov and A. I. Larkin, Phys. Lett. A **26**, 238 (1968).

[14] K. Maki, Prog. Theor. Phys. **39**, 897 (1968).

[15] R. S. Thompson, Phys. Rev. B **1**, 327 (1970).

[16] M. Yu. Reizer, Phys. Rev. B **45**, 12949 (1992).

[17] E. Abrahams and T. Tsuneto, Phys. Rev. **152**, 416 (1966).

[18] J. M. B. Lopes dos Santos and E. Abrahams, Phys. Rev. B **31**, 172 (1985).

[19] S. Hikami, A. I. Larkin, and Y. Nagaoka, Prog. Theor. Phys. **63**, 707 (1980).

[20] S. Kobayashi, A. Nakamura, and F. Komori, J. Phys. Soc. Jpn. **59**, 4219 (1990).

[21] R. B. Laibowitz, E. I. Alessandrini, and G. Deutscher, Phys. Rev. B **25**, 2965 (1982).

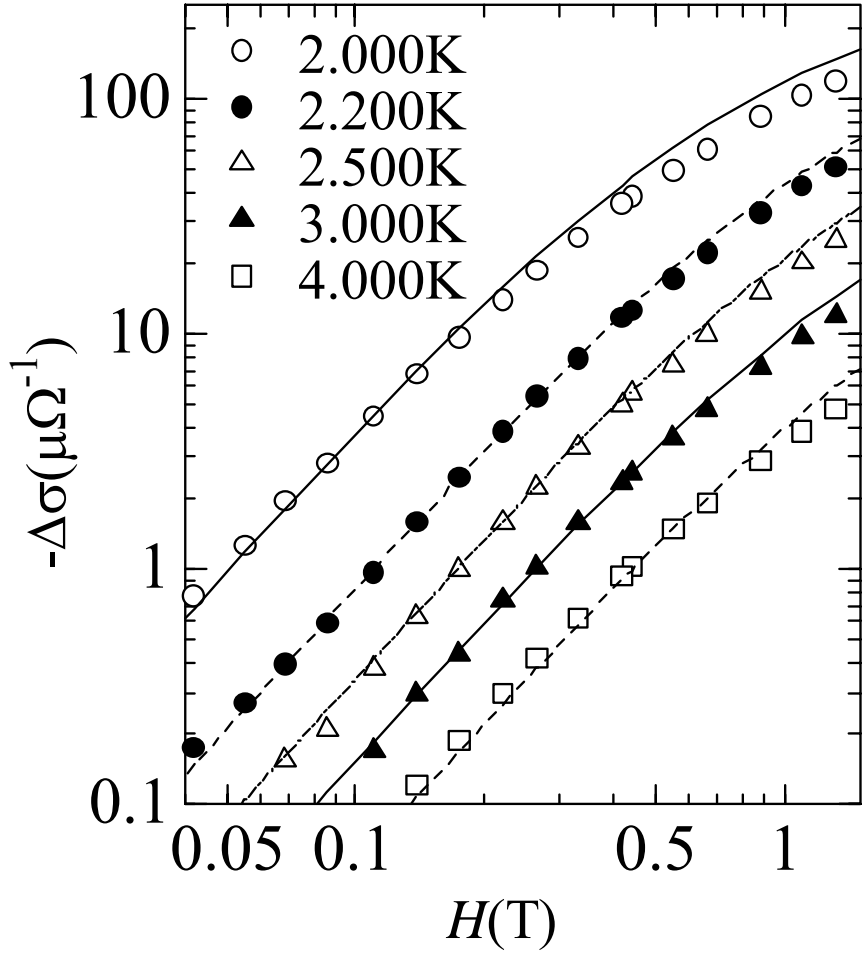

FIG. 1. The $H$ dependence of the magnetoconductance $\Delta\sigma$ of multi-layer film with $R_N = 3750\Omega$ at various temperatures. The lines are the best fit of the sum of Eqs.( 5 )-( 8 ) to the data using $D$ as a fitting parameter.

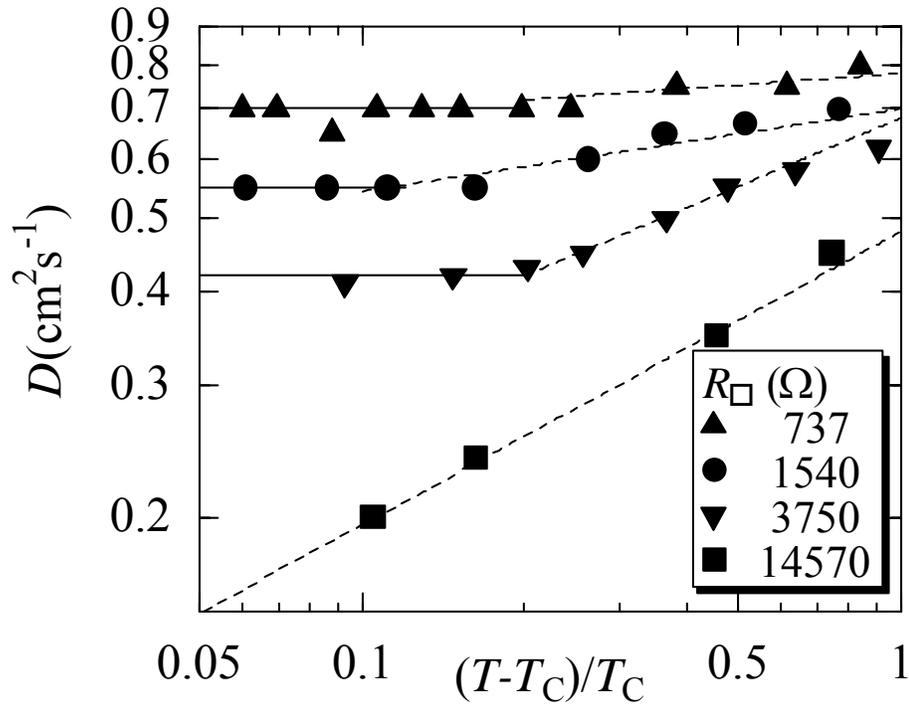

FIG. 2. The $(T-T_C)/T_C$ dependence of the diffusion constant $D$. The solid lines show the well-known homogenous behavior $\xi_P < \xi_S$. The broken lines are fitted by Eq.( 7 ) using θ as fitting parameter.

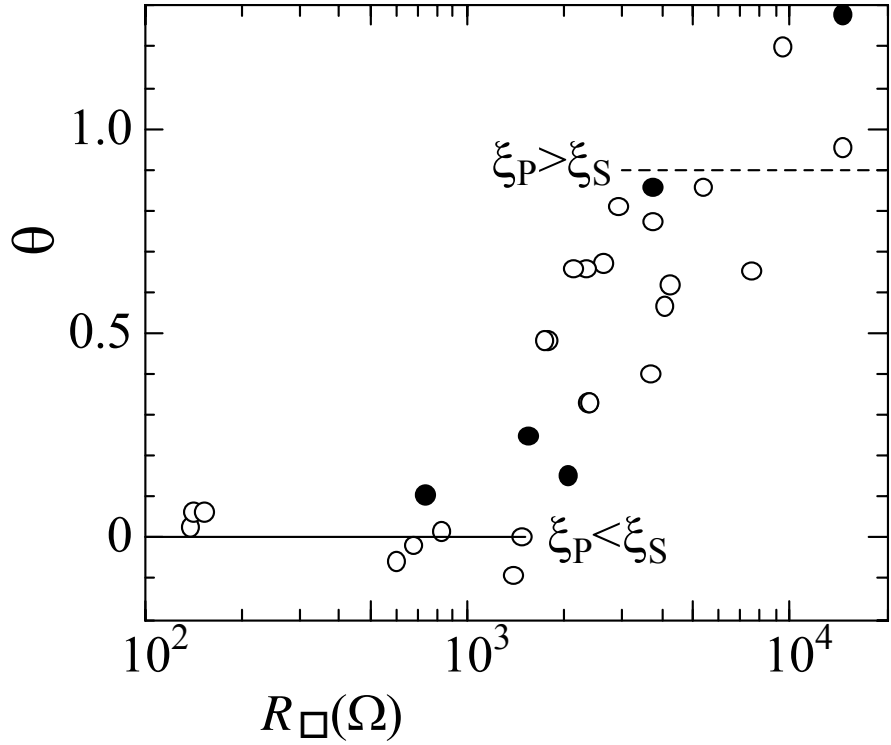

FIG. 3. The $R_\square$ dependence of $\theta$. The closed and open marks show $\theta$ obtained from the analysis of $\Delta\sigma$ and analyses of $H_{C2}(T)$ below $T_C$, respectively. The solid and dotted lines show the values of $\theta$ for homogenous and percolative films, respectively.

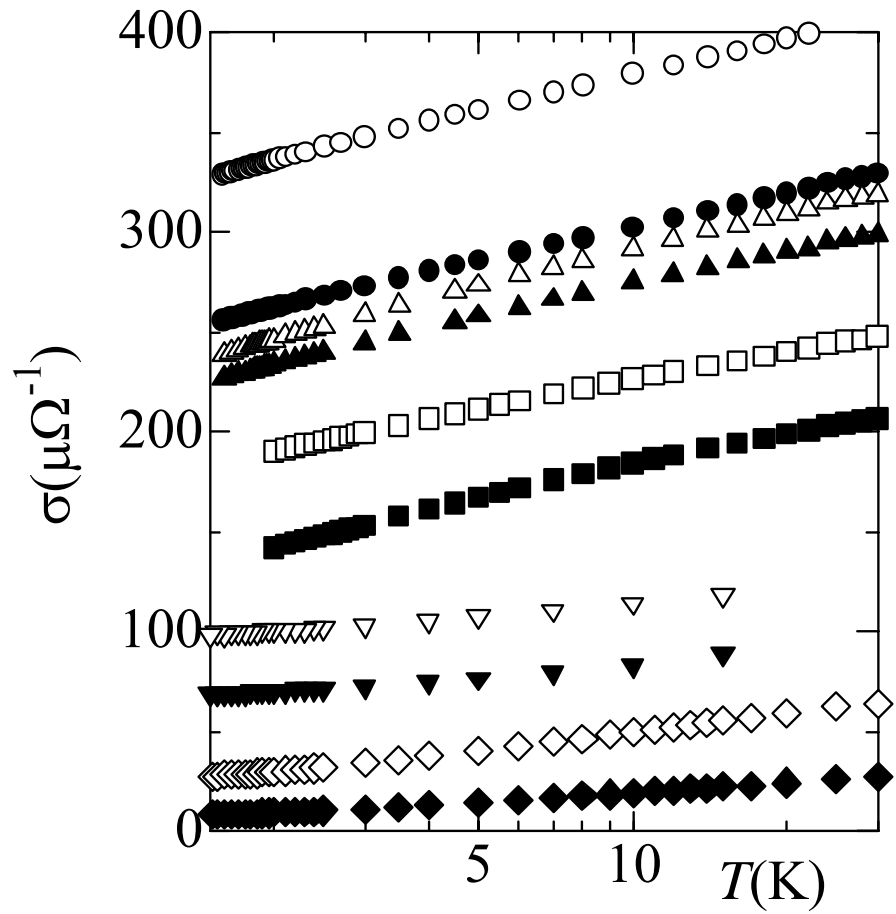

**FIG. 4. Temperature dependence of the sheet conductance for various films at $H$=5T. The value of $\sigma$ of the present films show the $\ln(T)$ dependence and indication of superconductivity cannot be observed.**

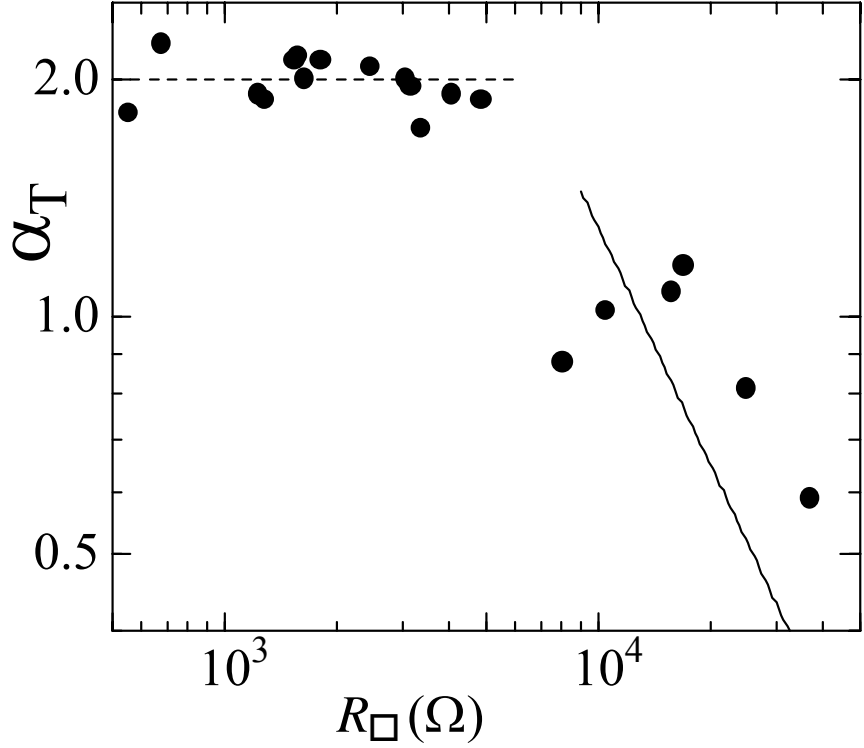

FIG. 5. The $R_\square$ dependence of $\alpha_T$. The value of horizontal axis $R_\square$ is defined at $T$=30K. Values of $\alpha_T$ were determined using Eq.( 1 ) from data in Fig. 4 at temperatures $10\,\text{K} < T < 30\,\text{K}$. The dotted line shows that $\alpha_T$ is independent of $R_\square$ for the case $\xi_P < L_{in}$. The solid line shows the relation $\alpha_T \propto 1/R_\square$, which is suggested from Eq.( 8 ).

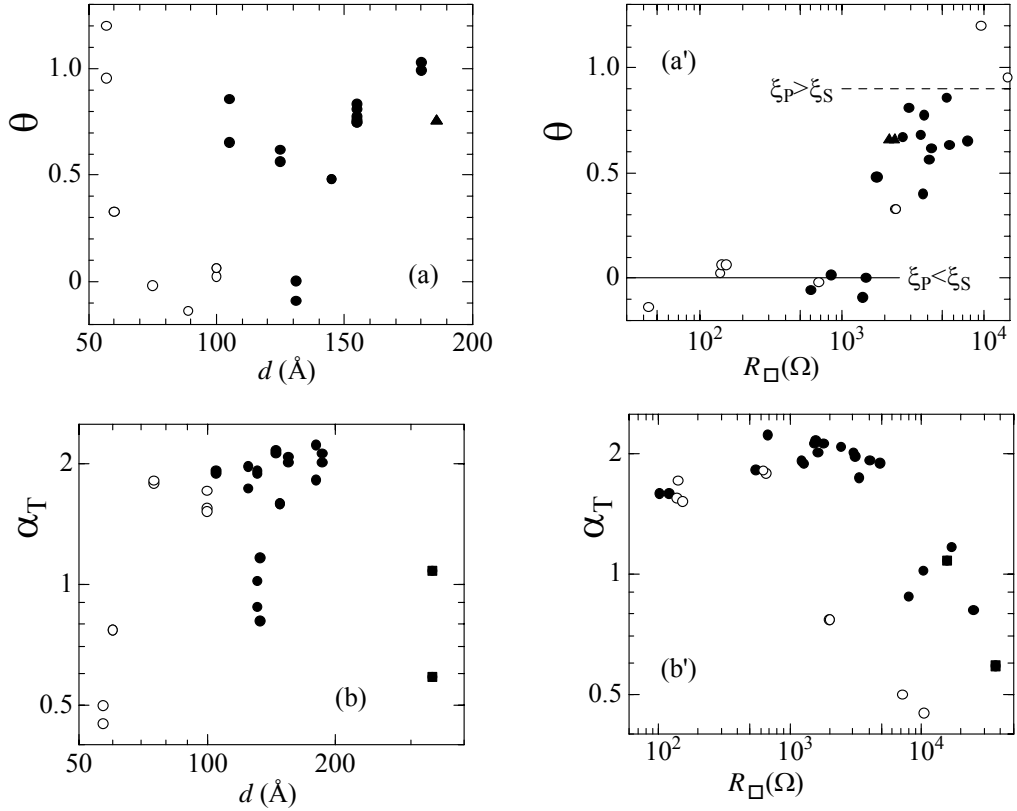

**FIG. 6.** *d* dependences of θ and $\alpha_T$ are shown in (a) and (b), respectively: $R_\square$ dependences of those are shown in (a') and (b'). The closed(●) and open(○) marks show the data of the multi-layer and single-layer films, respectively. The marks (▲) and (■) show a thick single-layer film and 10-layer films, respectively. In Fig. (b') the value of horizontal axis $R_\square$ is defined at *T*=30K.